# Superconducting Gap Anisotropy of LuNi$_2$B$_2$C Thin Films from Microwave Surface Impedance Measurements

A. Andreone, A. Cassinese, L. Gianni, M. Iavarone, F. Palomba, and R. Vaglio

*I.N.F.M. and Dipartimento di Scienze Fisiche, Universita' di Napoli "Federico II"*

*ITALY*

**Abstract.**

Surface impedance measurements of LuNi$_2$B$_2$C superconducting thin films as a function of temperature have been performed down to 1.5 K and at 20 GHz using a dielectric resonator technique. The magnetic penetration depth closely reproduces the standard B.C.S. result, but with a reduced value of the energy gap at low temperature. These data provide evidence for an anisotropic s-wave character of the order parameter symmetry in LuNi$_2$B$_2$C. From the evaluation of the real part of complex conductivity, we have observed constructive (type II) coherence effects in the electromagnetic absorption below $T_c$.

The rare earth R borocarbides with the generic formula $RNi_2B_2C$ are believed to be BCS-type superconductors. They have a moderately large density of states at the Fermi level, and it is generally agreed that the electron-phonon interaction is indeed the underlying mechanism for superconductivity [1].

The members of this family are either antiferromagnetic or nonmagnetic at low temperatures. Nonmagnetic compounds with R = Y and Lu exhibit fairly high $T_c$ values of about 15 - 16 K. Magnetism coexists with superconductivity for R = Dy, Ho, Er, and Tm, whereas only antiferromagnetic order occurs for R = Pr, Nd, Sm, Gd, and Tb. The antiferromagnetic ordering, and its competition (and even coexistence) with superconductivity, is conjectured to be driven by a nesting feature in the Fermi surface (FS) [2]. In the Lu compound, the *4f* band is fully occupied and therefore it is not magnetic. Since the *f* electrons occupy localized corelike states, the Fermi surface is expected to be similar to that of the other compounds.

A study of the Y and Lu compounds is therefore a prerequisite for the understanding of their magnetic counterparts, since they are free from the complications introduced by the presence of magnetism, and ideal for investigating the origin of superconductivity in borocarbides.

Even if, from the available experimental data, a dominating electron-phonon mediated mechanism seems to be a quite natural assumption, still there is a number of interesting properties that are not fully understood. Amongst them, the unusual temperature and field dependence of the electronic specific heat [3], the upward curvature of $H_{c2}(T)$ data close to $T_c$ [4], the observation of a quadratic flux line lattice at high fields [5].

During the years, these anomalies have been interpreted by different authors as hints for unconventional (d-wave) superconductivity [6]. Pair breaking effects caused by localized magnetic moments have been also invoked to explain the unusual temperature dependence of the upper critical field [7]. The presence of magnetic scattering is predicted to have an impact also on the low temperature behavior of penetration depth $\lambda$ of these compounds.

With the aim of shedding light on the origin of such anomalies, we report here on precise measurements of the microwave surface impedance of LuNi$_2$B$_2$C superconducting thin films, performed at 20 GHz by means of a sapphire dielectric resonator. Focus of this study is mainly on the temperature behavior of the magnetic penetration depth. It is well known in fact that λ(T) at low temperatures may be used as a very sensitive probe of the symmetry of the order parameter [8].

Very few microwave measurements have been performed since now on borocarbides, and only on the R = Y, Er, Tm, and Ho compounds of the family, leading to controversial results. Antiferromagnetic transitions at zero field were not always seen, but anomalies in the low-temperature surface impedance were clearly observed for ErNi$_2$B$_2$C, HoNi$_2$B$_2$C, and TmNi$_2$B$_2$C [9]. The temperature dependence of the penetration depth of YNi$_2$B$_2$C was found to disagree [9] or to be consistent with BCS expectations [10]. Results found for ErNi$_2$B$_2$C thin films supported a view where magnetic pair breaking affects the superconducting density of states [10].

LuNi$_2$B$_2$C thin films are prepared by a dc magnetron sputtering technique in ultra-high vacuum from a stoichiometric 5 cm diameter target prepared by arc-melting. Samples have been deposited on different substrates, with best results obtained using single crystal (100) MgO. In the following however we will report on films grown on single crystal sapphire 10×10×0.5 mm$^3$ substrates, because of their lowest microwave dielectric losses. Details of the procedure have been already reported elsewhere [11]. Briefly, deposition takes place starting from a chamber base pressure of 5 ÷ 6 · 10$^{-6}$ Pa and using Argon process gas at P$_{Ar}$ = 0.4 Pa. Substrates are placed on-axis at a distance of 9 cm from the target on a molybdenum block heater, held at a fixed temperature of about 800 °C. The sputtering rate is about 1 nm/s for a final thickness ranging between 300 and 500 nm. Four point resistivity probe indicates a resistivity *r* of the order of 100 μΩ or larger, a residual resistivity ratio *RRR = r(300 K)/r(T$_c$)* $\gg 3$, a critical temperature $T_c$ of about 15 K, $\Delta T_c$ values ranging between 0.4 and 0.8 K. Here $T_c$ is defined by the 90-10% criterion as the midpoint of the resistive transition, $\Delta T_c$ is the corresponding transition width. Inductive measurements show $T_c$ values slightly (about 0.5 K) lower, but with similar transition widths.

Structural characterization by a standard X-ray diffractometry shows a single phase growth with predominant c-axis orientation normal to the substrate plane. $\theta-2\theta$ spectra evidence also the presence of a small percentage of $Lu_2O_3$ oxide phase. Rocking curve measurement of the (002) peak gives a FWHM ~ 1.4º.

We have performed measurements of the magnetic penetration depth $\lambda$ and surface resistance $R_s$ as a function of temperature on two optimized samples. Each sample consists of two films, 300 nm thick, produced in the same deposition run, having therefore nominally the same structural, transport and superconducting properties. In the following we will show the results of measurements undertaken on the best of the two samples from $T_c$ down to 1.5 K. Data for the second sample follow a similar behavior.

The surface impedance $Z_s = R_s + jX_s = R_s + j\mu_0\omega\lambda$, where $\omega$ is the angular frequency, is investigated by means of a dielectrically loaded cylindrical resonator [12]. It consists of an OFHC (Oxygen Free High Conductivity) copper cylindrical shield of diameter 9.5 mm short-circuited at both ends by two superconducting films under test. A cylindrical dielectric sapphire rod of 7 mm diameter and 3.5 mm length is placed between the two parallel plates. The $TE_{011}$ field of the resonator is excited and detected by two semi-rigid coaxial cables, each having a small loop at the end, and the resonant frequency $f$ and Q-factor are measured in the transmission mode.

The effective (measured) surface impedance $Z_{seff} = R_{seff} + j\mu_0 \omega \lambda_{eff}$ is related to the surface impedance $Z_s$ of the "bulk" material in the full temperature range but very close to $T_c$ through the equation [13]:

$$Z_{s_{eff}} = R_s\left[\coth(t/\lambda) + \frac{t}{\lambda}\frac{1}{\sinh^2(t/\lambda)}\right] + j\mu_0 \omega \lambda \coth(t/\lambda)$$

(1)

where t is the sample thickness.

The effective surface resistance $R_{seff}$ is obtained measuring the quality factor of the resonator in unloaded condition $Q_u$, neglecting radiation and dielectric contribution, and using the formula:

$$R_{s_{eff}}(T) = \frac{\Gamma}{Q_u(T)} - \frac{\Gamma^2}{\Gamma + \Gamma_{lat}}\frac{1}{Q_{Cu}(T)}$$

(2)

Here $G$ is the geometrical factor associated with the field distribution on the film surface, whereas $G_{lat}$ takes into account contribution to losses due to the lateral copper walls. $Q_{Cu}$ is the unloaded quality factor of the sapphire resonator measured replacing the superconducting endplates with two OFHC copper foils. Below 50 K and at about 20 GHz, the surface impedance of copper is well described by the anomalous skin effect regime, therefore $Q_{Cu}$ can be considered as independent of temperature. For this geometry it is $G << G_{lat}$, so that the correction due to the lateral walls is within 15% in the overall measurement range.

The effective change in penetration depth $\Delta\lambda_{eff}$ is extracted using the relation:

$$\Delta\lambda_{eff}(T) \approx \beta(T)\left[\left(\frac{\Delta f(T)}{f(T_{min})}\right)_{Cu} - \left(\frac{\Delta f(T)}{f(T_{min})}\right)_{Sc}\right] + \Delta\delta_{Cu}(T)$$

(3)

where $\Delta f(T) = f(T) - f(0) \approx f(T) - f(T_{min})$ represents the change of the resonance frequency in the unperturbed (Cu) and perturbed (Sc) configuration, $\Delta\delta_{Cu}(T)$ represents the temperature variation of the copper skin depth, and $\beta(T) = \Gamma/[\pi \mu_0 f_{Sc}(T)]$.

Here the anomalous skin effect behavior implies $\Delta f_{Cu}(T)$, $\Delta\delta_{Cu} \approx 0$ at all temperatures of interest, so that

$$\Delta\lambda_{eff}(T) \approx \frac{\Gamma}{\pi\mu_0 f(T_{min})}\left[\left(\frac{f(T_{min})}{f(T)}\right)_{Sc} - 1\right]$$

(4)

In fig. 1 the penetration depth data, extracted from the resonant frequency measurements using eq. (4) and inverting eq. (1), are displayed as a function of temperature. The data are fitted to the B.C.S. theory, using the Halbritter code [14], shown as a continuous curve in the graph. The coherence length $\xi_0$, the critical temperature $T_c$ and the mean free path $\ell$ are set as input values, taken either from literature ($\xi_0$) or from transport measurements ($T_c$ and resistivity ρ), and the London penetration depth at zero temperature $\lambda_L(0)$ is extracted from the numerical fit. The strong coupling ratio $2\Delta(0)/k_B T_c$ is taken to be the B.C.S. value 3.5, consistently with previous findings [15]. All the parameters are reported in Table I. Data agree well with expectations for a weak-coupled s-wave superconductor, but at the lowest temperatures they show a significant departure. This is more clearly displayed in fig. 2, where the quantity $\Delta\lambda(T) = \lambda(T) - \lambda(0)$ is plotted as a function of temperature for $T/T_c < 0.5$. The B.C.S. curve using parameters taken from table I is represented by the dashed line. One can see that a good agreement between theory and data can be achieved only using a reduced value of the energy gap, $2\Delta(0)/k_B T_c = 2.0 \pm 0.2$ (continuous line). This is to our understanding a striking evidence for the existence of a strongly anisotropic s-wave gap in the $LuNi_2B_2C$ borocarbide. Indeed, in a band structure study Dugdale *et al.* [16] showed experimentally a rather complicated Fermi surface for this compound, revealing the presence of a sheet capable of nesting. The observed anisotropy was consistent with the observation of a square flux-line lattice and the unusual upper critical field behavior reported in previous experiments. Recent photoemission experiments [17] in the parent compound $YNi_2B_2C$ provided also spectroscopic evidence for an anisotropic s-wave gap.

It is worthwhile to mention, however, that in our case the gap anisotropy $D(0)_{max}/D(0)_{min} \approx$ 1.75, is more pronounced than results from other measurements: electronic Raman spectroscopy [18] for example shows a maximum anisotropy of 10%.

For completeness, we have also analyzed the penetration depth data taking into account the effect of correlation between magnetic impurities at low temperature, which gives rise to a spin-flip scattering frustration as T goes to zero [7]. This is accomplished introducing a temperature dependent magnetic scattering amplitude $G(T) = G_0(1+bT/q)/(1+T/q)$, where $G_0 = G(0)$, $q$ is a characteristic temperature describing the effects of correlation, $b$ is a parameter describing the relative change in $G$ caused by correlation. The presence of the temperature dependent $G$ has of course an impact on $\lambda(T)$. At low temperature one can write

$$\lambda(T) = \lambda(0)\left[\left(\frac{\pi}{2} - \alpha(\Gamma)\right)\Delta - \Gamma\left[\left(\frac{2}{3} - \frac{3}{4}sin\alpha(\Gamma)\right) - \frac{sin3\alpha(\Gamma)}{12}\right]\right]^{-1/2}$$

(5)

where $D$ is the order parameter and $a(G) = arcos(D/G)$ for $G > D$, $a(G) = 0$ otherwise.

Even in the low temperature region $D$ strongly depends on T, since the decrease in the effectiveness of pair-breaking translates in an increase of the pairing amplitude ("recovery" effect). The combined effect of $D(T)$ and $G(T)$ gives rise to an almost linear dependence of the magnetic penetration depth near T = 0 K, displayed as a dot-dashed line in fig. 2. Using the specific values of parameters *b*, *q* and $G_0$ taken from ref. [7], one can see that the model prediction can be hardly reconciled with the experimental data..

It has been recently proposed an interpretation of a number of unusual features presented by borocarbides in terms of a three-dimensional version of $d_{x^2-y^2}$ superconductivity. The experimentally observed saturation of the penetration depth at the lowest temperature however argues strongly against the possibility of nodes in the gap function of LuNi$_2$B$_2$C. For the sake of comparison, in fig. 2 we have also plotted the quadratic dependence expected for a dirty $d_{x^2-y^2}$ superconductor (dotted line) [19].

Using eq. (2) and inverting eq. (1), we can extract from the experiment the "bulk" surface resistance as a function of temperature. Assuming the usual $f^2$ scaling, the comparison of the measured $R_s$ with previous results on both single crystals and thin films of parent compounds evidences values which are between one and two orders of magnitude lower. Owing also to the good structural and surface quality of our samples, this makes us confident that the observed microwave dissipation is mainly intrinsic in origin. In spite of this, data cannot be consistently fitted in the overall temperature range within a B.C.S. framework using the same parameters shown in table I, since the level of losses is still too high. Surface impurities and other structural imperfections may of course affect the surface resistance in various ways. We speculate however that the observed losses may be an evidence of an unusual number of "unpaired" charge carriers arising from the same sheet in the FS which is responsible for the gap anisotropy. Indeed, for T less than $T_c/2$ the surface resistance behavior can be described using the standard B.C.S. exponential dependence:

$$R_s - R_{res} \approx A\sqrt{\Delta/k_B T}\, e^{-\Delta/k_B T}$$

(6)

where $R_{res}$ is the residual surface resistance and A is a phenomenological parameter. The value found for the ratio $2D(0)/k_BT_c$ lies between 1.5 and 1.8, depending mainly on the value chosen for the constant term $R_{res}$, in fair agreement therefore with penetration depth results at low temperatures. In fig. 3 the quantity $R_s - R_{res}$ is displayed, together with the low temperature B.C.S. fit ($2D(0)/k_BT_c = 1.6$).

In this picture losses can be ascribed to the existence of remnant scattering states due to the anisotropy in the energy gap or to the presence of strong electron-electron scattering in the nested regions of the FS [20].

From the surface resistance and magnetic penetration depth data, it is then possible to extract the temperature dependence of the complex conductivity $\sigma = \sigma_1 - j\sigma_2$, using the well-known equations:

$$\sigma_1 = \frac{2\mu_0 \omega R_s X_s}{(R_s^2 + X_s^2)^2} \quad (a); \qquad \sigma_2 = \frac{\mu_0 \omega (X_s^2 - R_s^2)^2}{(R_s^2 + X_s^2)^2} \quad (b)$$

(7)

Since $X_s \gg R_s$ in almost all the temperature range, $\sigma_2 \approx m_0 \omega/l^2$, retaining of course the B.C.S. functional dependence observed for the magnetic penetration depth. $\sigma_1$ instead first starts to increase below the critical temperature, reaching a broad maximum at about $0.8T_c$, and then decreases to a constant value at low temperatures. Subtracting in the analysis of the data the residual value $R_{res}$ found from the fit of surface resistance, one can force the real part of conductivity to reach a zero value at T = 0 K, which is equivalent to forcing the quasi-particles contribution to approach zero with decreasing temperature (fig.4). This procedure, however, doesn't change the non monotonous dependence displayed by the $\sigma_1$(T) data. The increase in the real part of conductivity below $T_c$ has to be related to the development of a singularity in the density of states at the gap edge, which leads therefore to the observation of a type II coherence (Hebel-Slichter) peak.

The experimental results are in good agreement with the dirty limit ($\ell \ll x_0$) B.C.S. weak coupling theory. Indeed, in fig. 4 we have compared the conductivity data with the B.C.S. expectation (dotted line) computed using a modified two-fluid model [21] and the standard $2D(0)/k_BT_c = 3.5$ ratio. In spite of the observed discrepancy at very high and very low temperature (the model loses accuracy close to $T_c$, and as T goes to zero in our data the energy gap is reduced), the agreement between the curve and measurements, in terms of position and intensity of the coherence peak, is remarkable.

Most NMR studies till now reported the absence of the Hebel-Slichter peak [22]. To our knowledge, to date there is only one report from a $^{13}$C NMR study of YNi$_2$B$_2$C polycrystalline samples where a small enhancement in the spin-lattice relaxation time temperature dependence just below $T_c$ has been observed [23].

In conclusion, data taken from surface impedance measurements of LuNi$_2$B$_2$C superconducting thin films show:

- the full consistency with a conventional s-wave phonon mediated framework. No evidence of d-wave symmetry has been found, nor magnetic impurities correlation effects have been proved to describe our experimental findings.

- the existence of a strong anisotropy in the energy gap, revealing itself in the low temperature dependence of both the real and imaginary part of $Z_s$. This is a strong support to a common view [16, 17, 24] that most unusual features observed in borocarbides may be related to the presence of deep minima in the gap function.

- the evidence of a clear peak in the $s_1$(T) dependence below $T_c$ to be associated to type II coherence effects in the electromagnetic absorption. This is a further direct confirmation that LuNi$_2$B$_2$C behaves as a conventional BCS s-wave superconductor.

Useful discussions with Prof. V. Z. Kresin and Dr. M. Salluzzo are gratefully acknowledged. The authors wish to thank A. Maggio and S. Marrazzo for technical support.

**Table captions**

Table I: B.C.S. fitting parameters for the experimental $l(T)$ data of a LuNi$_2$B$_2$C sample

| T$_c$ (K) | $\Delta(0)$ (meV) | $\lambda_L(0)$ (nm) | $x_0$ (nm) | $\ell^a$ (nm) |
|:---:|:---:|:---:|:---:|:---:|
| 14.2 | 2.1 | 275 ± 20 | 7 | 2 |

---

[a] from ref. [25]

**Figure captions**

Figure 1: Temperature dependence of the penetration depth $\lambda$ and fit to the B.C.S. theory (continuous line)

Figure 2: $\lambda$(T) - $\lambda$(0) as a function of the reduced temperature for T < $T_c$/2. Dashed and continuous lines represent the fit using the B.C.S. theory and $\Delta$(0) = 2.1 meV and 1.2 meV respectively, the dot-dashed line is the result assuming the presence of correlated magnetic impurities (see text), the dotted curve expresses the quadratic dependence expected from a dirty d-wave model

Figure 3: Temperature dependence of the B.C.S. contribution to the surface resistance $R_s$ - $R_{res}$ at 20 GHz, and fit to the B.C.S. theory (dotted line) at low temperature with a reduced gap value ($\Delta$(0) = 1.0 meV)

Figure 4: real part of the complex conductivity $\sigma_1$ as a function of temperature and fit to the B.C.S. weak-coupling dependence (dotted line)

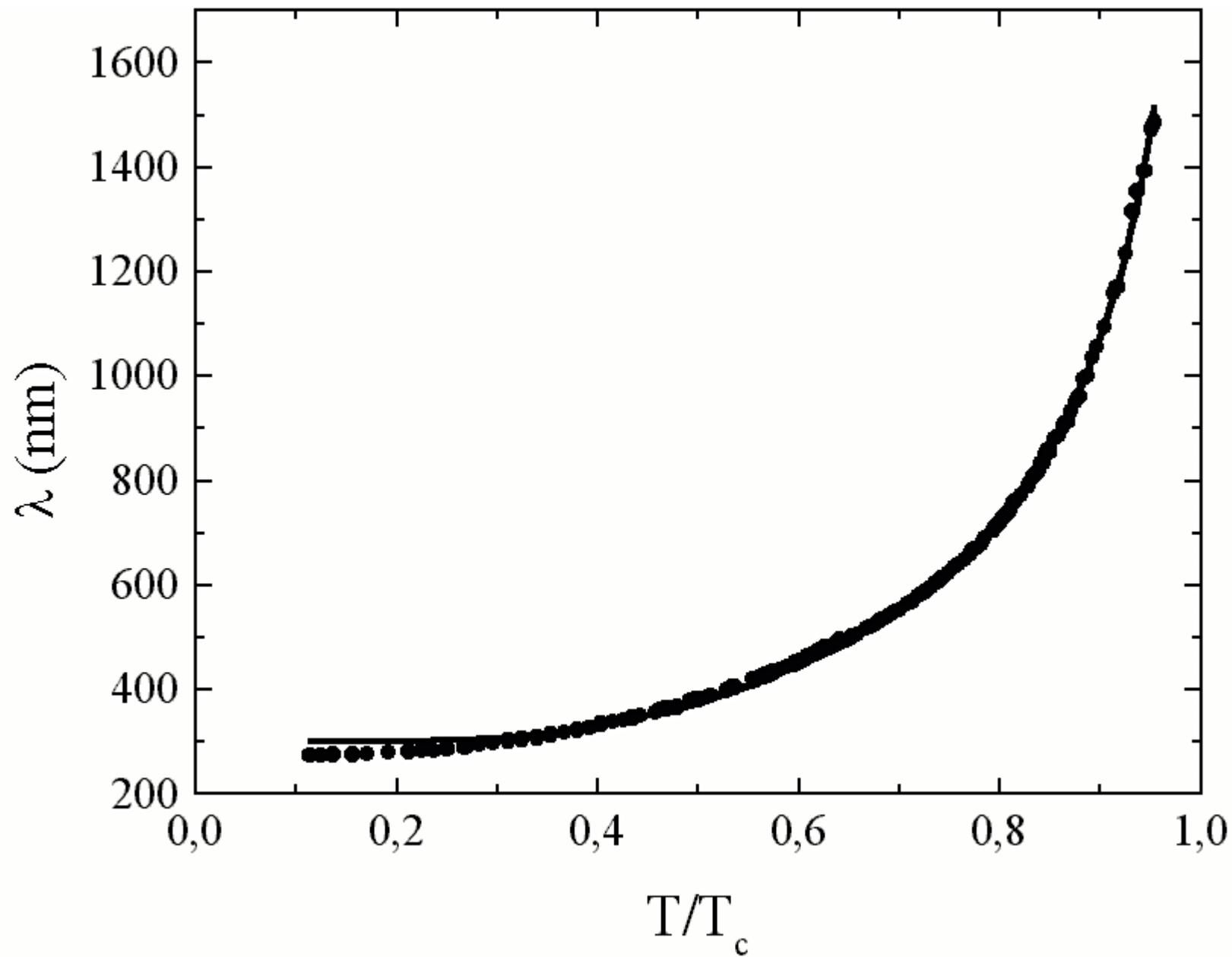

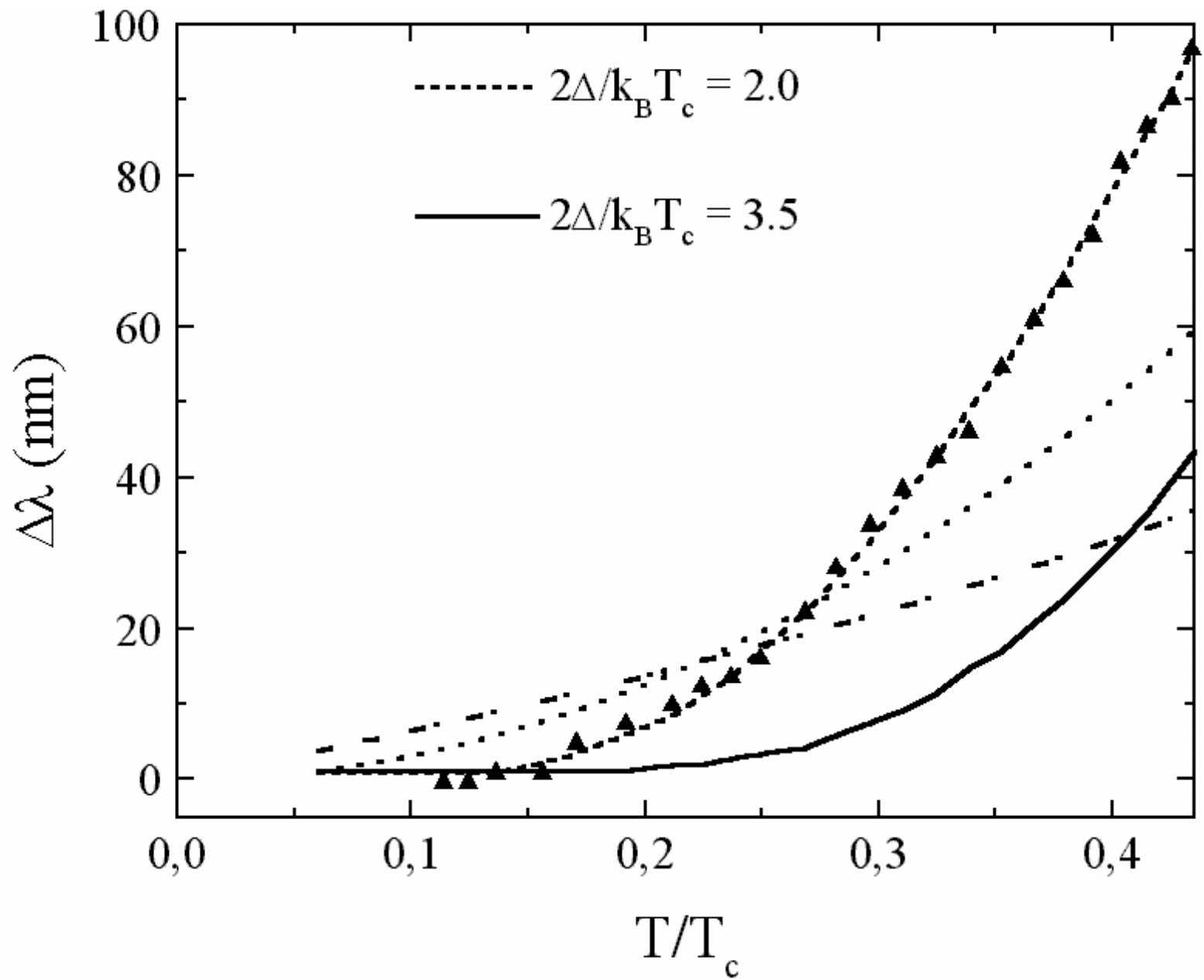

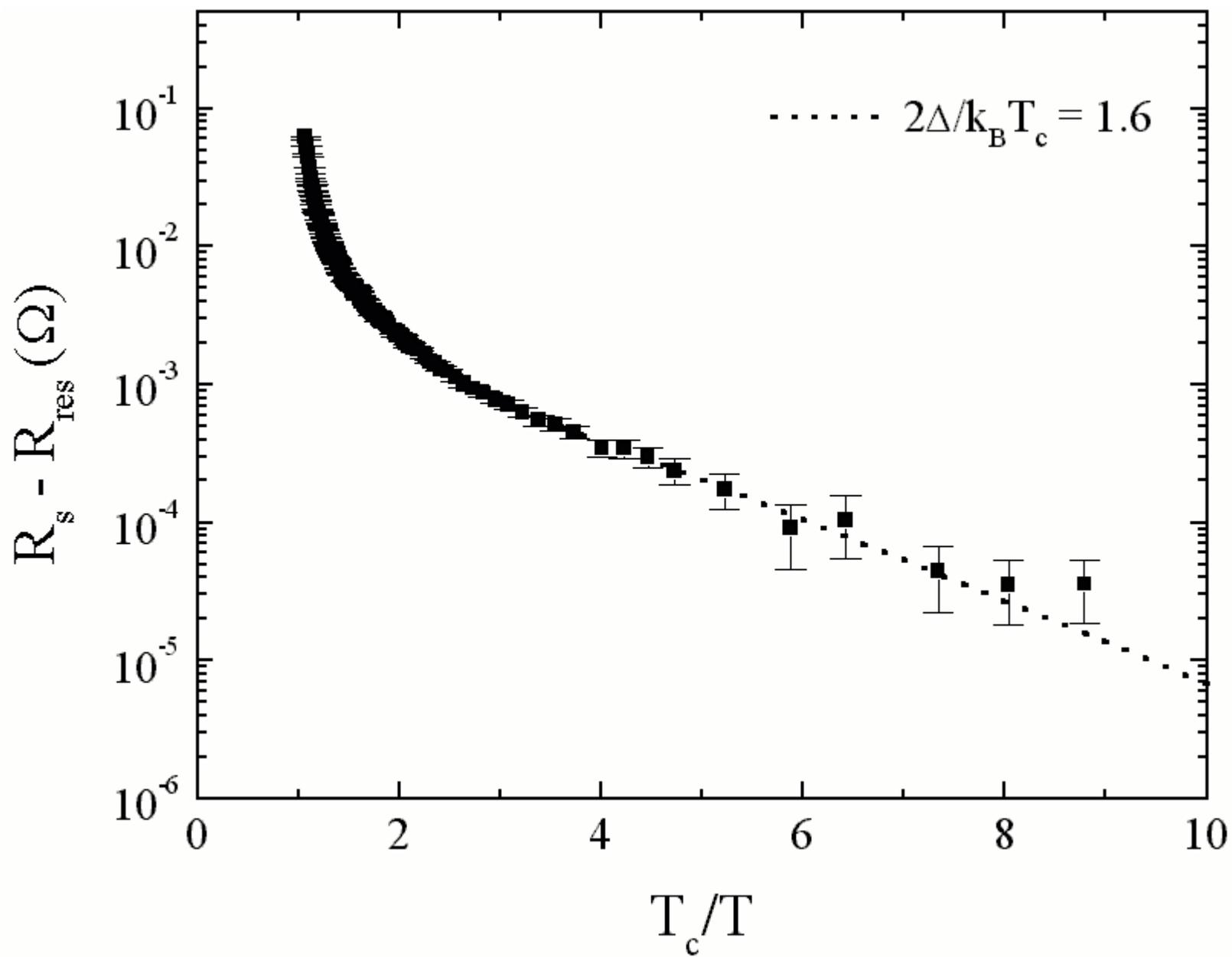

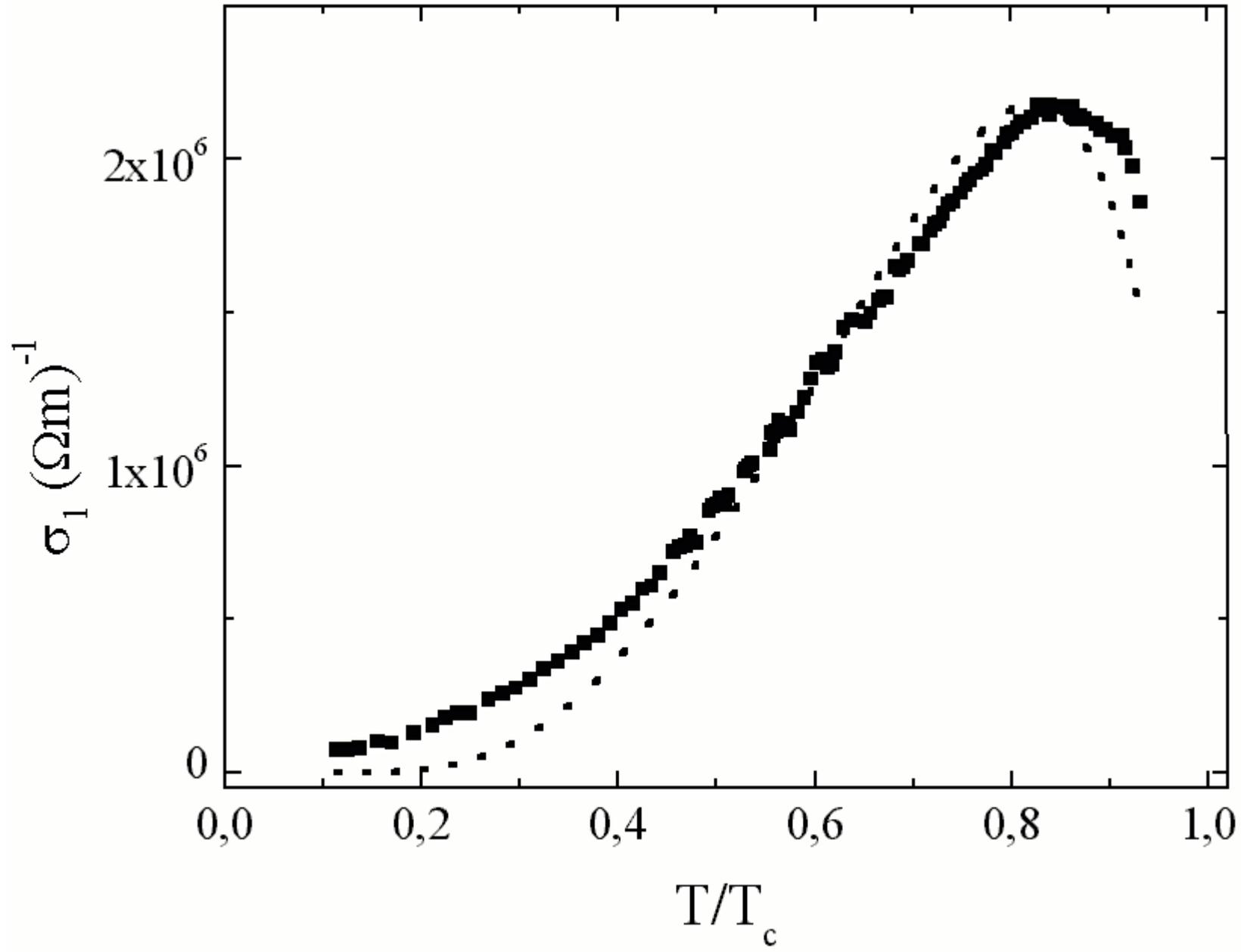